\documentclass[12pt]{iopart}

\usepackage{iopams}
\usepackage{amstext}
\usepackage{mathabx}
\usepackage{graphicx}
\usepackage{verbatim} 
\usepackage{color}

\def\be{\begin{equation}}
\def\ee{\end{equation}}
\def\bea{\begin{eqnarray}}
\def\eea{\end{eqnarray}}
\newcommand{\ket}[1]{\mbox{$|#1\rangle$}}
\newcommand{\bra}[1]{\mbox{$\langle#1|$}}
\newcommand{\avg}[1]{\mbox{$\langle#1\rangle$}}
\def\taud{\tau_{\footnotesize\textrm{delay}}}
\def\Gammaopt{\Gamma_{\footnotesize\textrm{opt}}}
\def\Pnoise{P_{\footnotesize\textrm{noise}}}
\def\keff{k_{\footnotesize\textrm{eff}}}
\def\kappaex{\kappa_{\footnotesize\textrm{ex}}}
\def\kappain{\kappa_{\footnotesize\textrm{in}}}
\newcommand{\opdagger}[2]{\mbox{$\hat{#1}_{#2}^{\dagger}$}}
\newcommand{\op}[2]{\mbox{$\hat{#1}_{#2}$}}

\newcommand{\m}[1]{\mbox{$\mathbf{#1}$}}

\begin{document}

\title[Slowing and stopping light using an optomechanical crystal array]{Slowing and stopping light using an optomechanical crystal array}

\author{D.E. Chang$^{1,4}$,\footnote[0]{$^4$  These authors contributed equally to this work.} A.H. Safavi-Naeini$^{2,4}$, M. Hafezi$^3$, O. Painter$^{2}$}

\address{$^1$Institute for
Quantum Information and Center for the Physics of Information,
California Institute of Technology, Pasadena, CA 91125}
\address{$^2$Thomas J. Watson, Sr., Laboratory of Applied Physics,
California Institute of Technology, Pasadena, CA 91125}
\address{$^3$Joint Quantum Institute and Department of Physics, University of Maryland, College Park, MD 20742}
\ead{opainter@caltech.edu}

\date{\today}

\begin{abstract}
One of the major advances needed to realize all-optical
information processing of light is the ability to delay or
coherently store and retrieve optical information in a rapidly
tunable manner. In the classical domain, this optical buffering is
expected to be a key ingredient to managing the flow of
information over complex optical networks. Such a system also has
profound implications for quantum information processing, serving
as a long-term memory that can store the full quantum information
contained in an optical pulse. Here we suggest a novel approach to
light storage involving an optical waveguide coupled to an
optomechanical crystal array, where light in the waveguide can be
dynamically and coherently transferred into long-lived mechanical
vibrations of the array. Under realistic conditions, this system
is capable of achieving large bandwidths and storage/delay times
in a compact, on-chip platform.
\end{abstract}

\pacs{37.10.Vz,42.50.Pq, 42.50.Ar, 42.50.Lc, 42.79-e}

\maketitle

\section{Introduction}

Light is a natural candidate to transmit information across large
networks due to its high speed and low propagation losses.  A
major obstacle to building more advanced optical networks is the
lack of an all-optically controlled device that can robustly delay
or store optical wave-packets over a tunable amount of time.  In
the classical domain, such a device would enable all-optical
buffering and switching, bypassing the need to convert an optical
pulse to an electronic signal. In the quantum realm, such a device
could serve as a memory to store the full quantum information
contained in a light pulse until it can be passed to a processing
node at some later time.

A number of schemes to coherently delay and store optical
information are being actively explored.  These range from tunable
coupled resonator optical waveguide~(CROW)
structures~\cite{yanik04,scheuer05}, where the propagation of
light is dynamically altered by modulating the refractive index of
the system, to electromagnetically induced transparency~(EIT) in
atomic media~\cite{fleischhauer00,fleischhauer05}, where the
optical pulse is reversibly mapped into internal atomic degrees of
freedom. While these schemes have been demonstrated in a number of
remarkable experiments~\cite{liu01,phillips01,okawachi06,xu07},
they remain difficult to implement in a practical setting.  Here,
we present a novel approach to store or stop an optical pulse
propagating through a waveguide, wherein coupling between the
waveguide and a nearby nano-mechanical resonator array enables one
to map the optical field into long-lived mechanical excitations.
This process is completely quantum coherent and allows the delay
and release of pulses to be rapidly and all-optically tuned. Our
scheme combines many of the best attributes of previously proposed
approaches, in that it simultaneously allows for large bandwidths
of operation, on-chip integration, relatively long delay/storage
times, and ease of external control.  Beyond light storage, this
work opens up the intriguing possibility of a platform for quantum
or classical all-optical information processing using mechanical
systems.

\section{Description of system: an optomechanical crystal array}

An optomechanical crystal~\cite{eichenfield09} is a periodic
structure that constitutes both a photonic~\cite{joannopoulos08}
and a phononic~\cite{maldovan06} crystal. The ability to engineer
optical and mechanical properties in the same structure should
enable unprecedented control over light-matter interactions.
Planar two-dimensional (2D) photonic crystals, formed from
patterned thin dielectric films on the surface of a microchip,
have been succesfully employed as nanoscale optical circuits
capable of efficiently routing, diffracting, and trapping light.
Fabrication techniques for such 2D photonic crystals have matured
significantly over the last decade, with experiments on a Si
chip~\cite{notomi08} demonstrating excellent optical transmission
through long~($N>100$) linear arrays of coupled photonic crystal
cavities.  In a similar Si chip platform it has recently been
shown that suitably designed photonic crystal cavities also
contain localized acoustic resonances which are strongly coupled
to the optical field via radiation pressure~\cite{eichenfield09}.
These planar optomechanical crystals~(OMCs) are thus a natural
candidate for implementation of our proposed slow-light scheme.
\begin{figure}[htbp]
\begin{center}
\includegraphics[width=17cm]{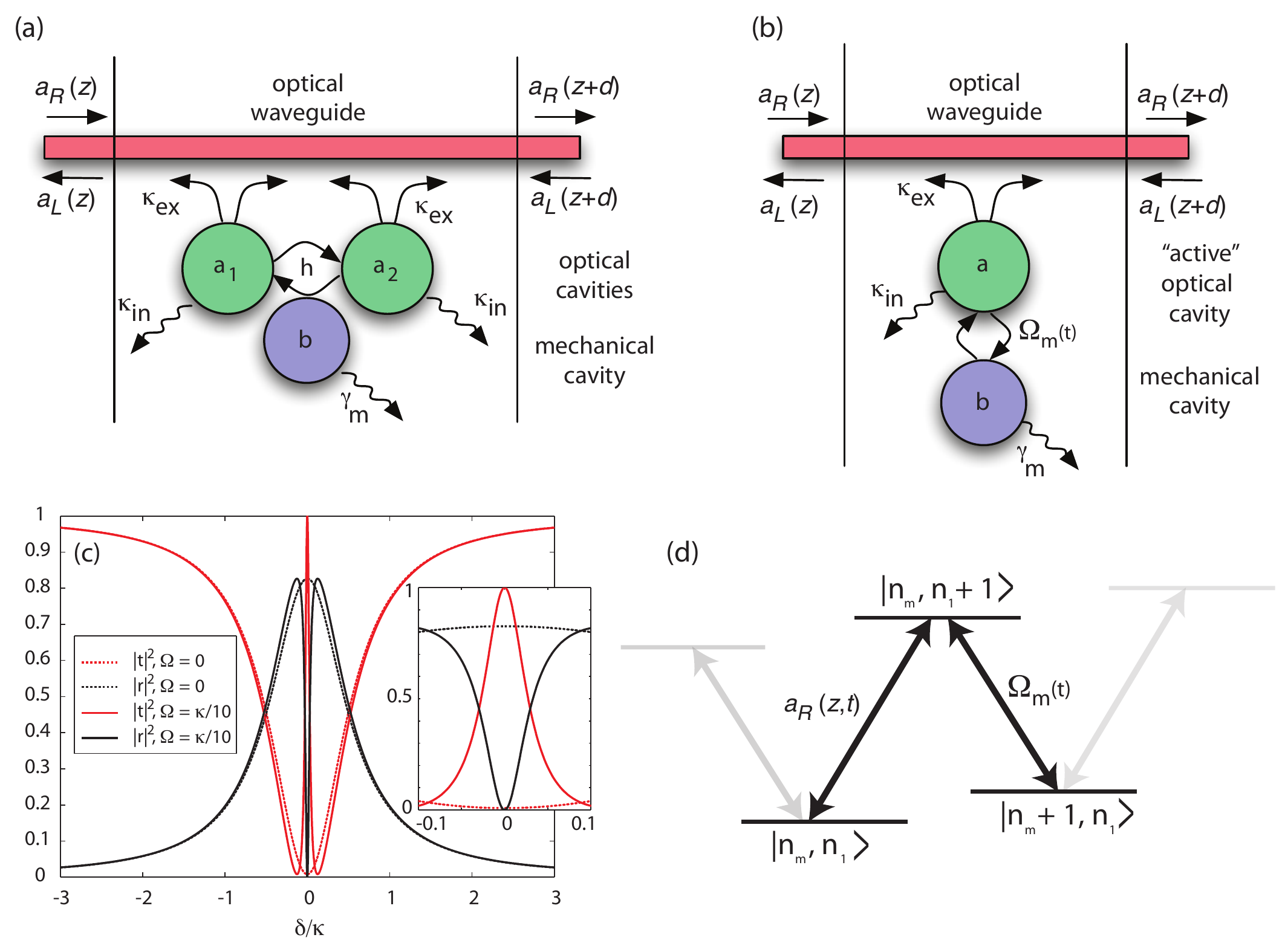}
\end{center}
\caption{(a) Illustration of a double optical cavity system forming the unit cell of the optomechanical array. A two-way
  optical waveguide is coupled to a pair of optical cavity modes $a_1$ and $a_2$, whose resonance frequencies differ by
  the frequency of the mechanical mode $b$. Both optical modes leak energy into the waveguide at a rate $\kappaex$ and
  have an inherent decay rate $\kappain$. The mechanical resonator optomechanically couples the two optical resonances
  with a cross coupling rate of $h$. (b)A simplified system diagram where the classically driven cavity mode $a_2$ is
  effectively eliminated to yield an optomechanical driving amplitude $\Omega_m$ between the mechanical mode and cavity
  mode $a_1$. (c) Frequency-dependent reflectance~(black curve) and transmittance~(red) of a single array element, in
  the case of no optomechanical driving amplitude $\Omega_m=0$~(dotted line) and an amplitude of
  $\Omega_{m}=\kappaex/10$~(solid line). The inherent cavity decay is chosen to be $\kappain=0.1\kappaex$. (inset) The
  optomechanical coupling creates a transparency window of width ${\sim}4\Omega^2_m/\kappaex$ for a single element and
  enables perfect transmission on resonance, $\delta_k=0$.  (d) Energy level structure of simplified system. The number
  of photons and phonons are denoted by $n_1$ and $n_m$, respectively. The optomechanical driving amplitude $\Omega_m$
  couples states $\ket{n_m+1,n_1}\leftrightarrow\ket{n_m,n_1+1}$ while the light in the waveguide couples states
  $\ket{n_m,n_1}\leftrightarrow\ket{n_m,n_1+1}$. The two couplings create a set of $\Lambda$-type transitions analogous
  to that in EIT.\label{fig:system}}
\end{figure}

In the following we consider an optomechanical crystal containing
a periodic array of such defect cavities~(see
figures~\ref{fig:system}(a),(b)). Each element of the array contains two
optical cavity modes~(denoted $1,2$) and a co-localized mechanical
resonance. The Hamiltonian describing the dynamics of a single
element is of the form
\be
\tilde{H}_{\footnotesize\textrm{om}}=\hbar\omega_{1}\opdagger{a}{1}\op{a}{1}+\hbar\omega_{2}\opdagger{a}{2}\op{a}{2}+\hbar\omega_{m}\hat{b}^{\dagger}\hat{b}+{\hbar}h\left(\hat{b}+\hat{b}^{\dagger}\right)\left(\opdagger{a}{1}\op{a}{2}+\opdagger{a}{2}\op{a}{1}\right).\label{eq:H}
\ee
Here $\omega_{1,2}$ are the resonance frequencies of the two
optical modes, $\omega_{m}$ is the mechanical resonance frequency,
and $\op{a}{1},\op{a}{2},\hat{b}$ are annihilation operators for
these modes. The optomechanical interaction cross-couples the
cavity modes $1$ and $2$ with a strength characterized by $h$ and
that depends linearly on the mechanical displacement
$\hat{x}\propto(\hat{b}+\hat{b}^{\dagger})$. While we formally
treat $\op{a}{1},\op{a}{2},\hat{b}$ as quantum mechanical
operators, for the most part it also suffices to treat these terms
as dimensionless classical quantities describing the
positive-frequency components of the optical fields and mechanical
position. In addition to the optomechanical interaction described
by equation~(\ref{eq:H}), the cavity modes $1$ are coupled to a common
two-way waveguide~(described below). Each element is decoupled
from the others except through the waveguide.

The design considerations necessary to achieve such a system are
discussed in detail in the section ``Optomechanical crystal
design.'' For now, we take as typical parameters
$\omega_{1}/2\pi=200$~THz, $\omega_{m}/2\pi=10$~GHz,
$h/2\pi=0.35$~MHz, and mechanical and (unloaded) optical quality
factors of $Q_{m}\equiv\omega_{m}/\gamma_{m}{\sim}10^3$~(room
temperature)-$10^5$~(low temperature) and
$Q_{1}\equiv\omega_{1}/\kappa_{1,in}=3{\times}10^6$, where
$\gamma_{m}$ is the mechanical decay rate and $\kappa_{1,in}$ is
the intrinsic optical cavity decay rate. Similar
parameters have been experimentally observed in other OMC
systems~\cite{eichenfield09,safavi-naeini10b}. In practice, one
can also over-couple cavity mode $1$ to the waveguide, with a
waveguide-induced optical decay rate $\kappaex$ that is much
larger than $\kappain$.

For the purpose of slowing light, the cavity modes $2$ will be
resonantly driven by an external laser, so that to good
approximation $\op{a}{2}{\approx}\alpha_{2}(t)e^{-i\omega_{2}t}$
can be replaced by its mean-field value.  We furthermore consider
the case where the frequencies are tuned such that
$\omega_{1}=\omega_{2}+\omega_m$.  Keeping only the resonant terms
in the optomechanical interaction, we arrive at a simplified
Hamiltonian for a single array element~(see
figure~\ref{fig:system}(b)),
\be
H_{\footnotesize\textrm{om}}=\hbar\omega_{1}\opdagger{a}{1}\op{a}{1}+\hbar\omega_{m}\hat{b}^{\dagger}\hat{b}+\hbar\Omega_{m}(t)\left(\opdagger{a}{1}\hat{b}e^{-i(\omega_{1}-\omega_m)t}+h.c.\right).\label{eq:Hlinear}
\ee
Here we have defined an effective optomechanical driving amplitude
$\Omega_{m}(t)=h\alpha_{2}(t)$ and assume that $\alpha_{2}(t)$ is
real. Mode $2$ thus serves as a ``tuning'' cavity that mediates
population transfer~(Rabi oscillations) between the ``active''
cavity mode $1$ and the mechanical resonator at a controllable
rate $\Omega(t)$, which is the key mechanism for our stopped-light protocol. In the following analysis, we will focus exclusively on the active cavity mode and drop the
``$1$'' subscript.

A Hamiltonian of the form~(\ref{eq:Hlinear}) also describes an
optomechanical system with a single optical mode, when the cavity
is driven off resonance at frequency $\omega_1-\omega_m$ and
$\op{a}{1}$ corresponds to the sidebands generated
at frequencies ${\pm}\omega_m$ around the classical driving
field. For a single system, this Hamiltonian leads to efficient
optical cooling of the mechanical
motion~\cite{wilson-rae07,marquardt07}, a technique being used to
cool nano-mechanical systems toward their quantum ground
states~\cite{arcizet06,gigan06,schliesser06,cleland09}.
While the majority of such work focuses on how
optical fields affect the mechanical dynamics, here we show that
the optomechanical interaction strongly modifies optical field
propagation to yield the slow/stopped light phenomenon.
Equation~(\ref{eq:Hlinear}) is quite general and thus this
phenomenon could in principle be observed in any array of
optomechanical systems coupled to a waveguide. In practice, there
are several considerations that make the 2D OMC ``ideal.'' First,
our system exhibits an extremely large optomechanical coupling $h$
and contains a second optical tuning cavity that can be driven
resonantly, which enables large driving amplitudes $\Omega_m$
using reasonable input power~\cite{safavi-naeini10a}. Using two
different cavities also potentially allows for greater versatility
and addressability of our system. For instance, in our proposed
design the photons in cavity $1$ are spatially filtered from those
in cavity 2~\cite{safavi-naeini10a}. Second, the 2D OMC is an
easily scalable and compact platform. Finally, as described below,
the high mechanical frequency of our device compared to typical
optomechanical systems allows for a good balance between long
storage times and suppression of noise processes.

\section{Slowing and stopping light}
\subsection{Static regime}

We first analyze propagation in the waveguide when
$\Omega_{m}(t)=\Omega_m$ is static during the transit interval of
the signal pulse. As shown in the Appendix,
the evolution equations in a rotating frame for a single element
located at position $z_j$ along the waveguide are given by
\bea \frac{d\hat{a}}{dt} & = &
-\frac{\kappa}{2}\hat{a}+i\Omega_{m}\hat{b}+i\sqrt{\frac{c\kappaex}{2}}\left(\op{a}{R,\footnotesize\textrm{in}}(z_j)+\op{a}{L,\footnotesize\textrm{in}}(z_j)\right)+\sqrt{c\kappain}\op{a}{\footnotesize\textrm{N}}(z_j),\label{eq:dadt}
\\ \frac{d\hat{b}}{dt} & = &
-\frac{\gamma_m}{2}\hat{b}+i\Omega_{m}\hat{a}+\op{F}{N}(t).\label{eq:dbdt}
\eea
Equation~(\ref{eq:dadt}) is a standard input relation characterizing
the coupling of right- ($\op{a}{R,\footnotesize\textrm{in}}$) and left-propagating ($\op{a}{L,\footnotesize\textrm{in}}$) optical input fields in the
waveguide with the cavity mode. Here $\kappa=\kappaex+\kappain$ is
the total optical cavity decay rate, $\op{a}{\footnotesize\textrm{N}}(z)$
is quantum noise associated with the inherent optical cavity loss, and for simplicity we have assumed a linear dispersion relation $\omega_k=c|k|$ in the waveguide. Equation~(\ref{eq:dbdt})
describes the optically driven mechanical motion, which decays at
a rate $\gamma_m$ and is subject to thermal noise $\op{F}{N}(t)$.
The cavity mode couples to the right-propagating field through the
equation
\be
\left(\frac{1}{c}\frac{\partial}{{\partial}t}+\frac{\partial}{{\partial}z}\right)\op{a}{R}(z,t)=i\sqrt{\frac{\kappaex}{2c}}\delta(z-z_j)\hat{a}+ik_{0}\op{a}{R},\label{eq:waveeqn}
\ee
where $k_0=\omega_1/c$. We solve the above equations to find the
reflection and transmission coefficients $r,t$ of a single element
for a right-propagating incoming field of frequency $\omega_k$~(see Appendix). In the
limit where $\gamma_m=0$, and defining
$\delta_k{\equiv}\omega_k-\omega_1$,
\be r(\delta_k)=-\frac{\delta_{k}\kappaex}{\delta_k(-2i\delta_k+\kappa)+2i\Omega_m^2}, \ee
while $t=1+r$. Example reflectance and transmittance curves are
plotted in figure~\ref{fig:system}(c). For any non-zero $\Omega_m$, a
single element is perfectly transmitting on resonance, whereas for
$\Omega_m=0$ resonant transmission past the cavity is blocked.
When $\Omega_m{\neq}0$, excitation of the cavity mode is inhibited
through destructive interference between the incoming field and
the optomechanical coupling. In EIT, a similar effect occurs via
interference between two electronic transitions. This analogy is
further elucidated by considering the level structure of our
optomechanical system~(figure~\ref{fig:system}(d)), where the
interference pathways and the ``$\Lambda$''-type transition
reminiscent of EIT are clearly visible. The interference is
accompanied by a steep phase variation in the transmitted field
around resonance, which can result in a slow group velocity. These
steep features and their similarity to EIT in a single
optomechanical system have been
theoretically~\cite{agarwal09,schliesser10} and
experimentally studied~\cite{weis10,safavi-naeini10c}, while interference effects
between a single cavity mode and two mechanical modes have also
been observed~\cite{lin09}.
\begin{figure}[htbp]
\begin{center}
\includegraphics[width=17cm]{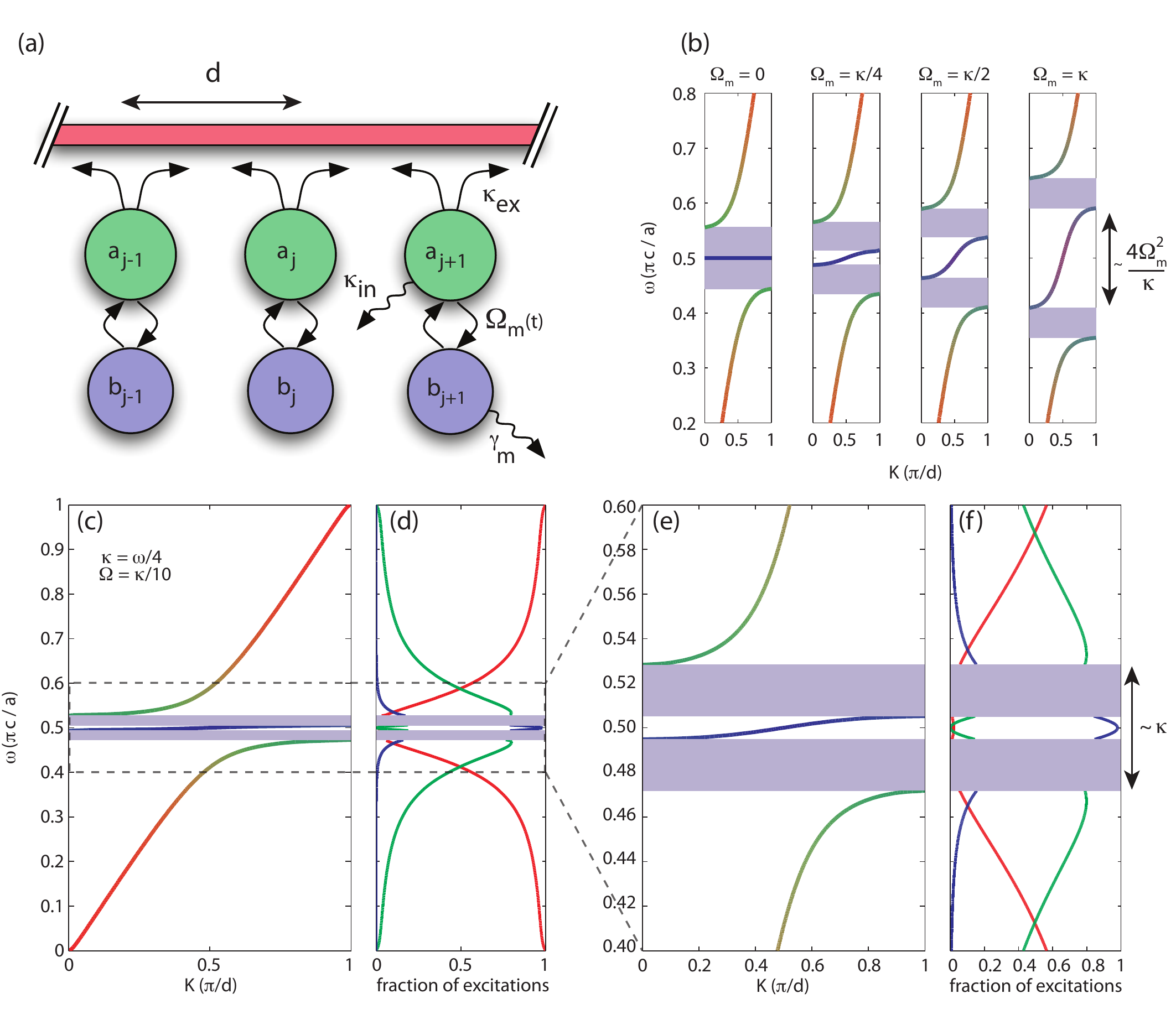}
\end{center}
\caption{(a) Illustration of an optomechanical crystal array. A
two-way optical waveguide is coupled to a periodic array of
optomechanical elements spaced by a distance $d$. The optical
cavity modes $a_j$ of each element leak energy into the waveguide
at a rate $\kappaex$ and have an inherent decay rate $\kappain$.
The mechanical resonator of each element has frequency $\omega_m$
and is optomechanically coupled to the cavity mode through a
tuning cavity~(shown in figure~\ref{fig:system}) with strength
$\Omega_m$. (b) The band structure of the system, for a range of
driving strengths between $\Omega_m=0$ and $\Omega_m=\kappa$. The
blue shaded regions indicate band gaps, while the color of the
bands elucidates the fractional occupation~(red for energy in the
optical waveguide, green for the optical cavity, and blue for
mechanical excitations). The dynamic compression of the bandwidth
is clearly visible as $\Omega_m\rightarrow0$. (c) Band structure
for the case $\Omega_m=\kappa/10$ is shown in greater detail. (d)
The fractional occupation for each band in (c) is plotted
separately. It can be seen that the polaritonic slow-light band is
mostly mechanical in nature, with a small mixing with the
waveguide modes and negligible mixing with the optical cavity
mode. Zoom-ins of figures (c) and (d) are shown in (e) and (f).
 \label{fig:bandstructure}}
\end{figure}
From $r,t$ for a single element, the propagation characteristics
through an infinite array~(figure~\ref{fig:bandstructure}(a)) can be
readily obtained via band structure
calculations~\cite{joannopoulos08}. To maximize the propagation
bandwidth of the system, we choose the spacing $d$ between
elements such that $k_{0}d=(2n+1)\pi/2$ where $n$ is a
non-negative integer.  With this choice of phasing the reflections from multiple elements
destructively interfere under optomechanical driving. Typical band structures are illustrated
in figures~\ref{fig:bandstructure}(b)-(f). The color coding of the
dispersion curves~(red for waveguide, green for optical cavity, blue for
mechanical resonance) indicates the distribution of energy or fractional occupation in the various degrees of freedom of the system in steady-state. Far away from the cavity resonance, the
dispersion relation is nearly linear and simply reflects the
character of the input optical waveguide, while the propagation is
strongly modified near resonance~($\omega=\omega_{1}=\omega_{2}+\omega_{m}$). In the absence of optomechanical
coupling~($\Omega_m=0$), a transmission band gap of width $\sim \kappa$ forms around the optical cavity
resonance (reflections from the bare optical cavity elements constructively interfere). In the
presence of optomechanical driving, the band gap splits in two (blue shaded regions) and a new
propagation band centered around the cavity resonance
appears in the middle of the band gap. For weak driving~($\Omega_m{\lesssim}\kappa$) the width of this band is ${\sim}4\Omega_m^2/\kappa$, while for strong driving~($\Omega_m{\gtrsim}\kappa$) one recovers the ``normal mode splitting'' of width ${\sim}2\Omega_m$~\cite{groblacher09a}. This relatively flat polaritonic band yields the
slow-light propagation of interest. Indeed, for small $\Omega_m$
the steady-state energy in this band is almost completely mechanical in character,
indicating the strong mixing and conversion of energy in the
waveguide to mechanical excitations along the array.

It can be shown that the Bloch wavevector near resonance is given
by~(see Appendix)
\be \keff{\approx}k_0+\frac{\kappaex\delta_k}{2d\Omega_m^2}+\frac{i\kappaex\kappain\delta_k^2}{4d\Omega_m^4}+\frac{(2\kappaex^3-3\kappaex\kappain^2+12\kappaex\Omega_m^2)\delta_k^3}{24d\Omega_m^6}.\label{eq:keff} \ee
The group velocity on resonance,
$v_g=(d\keff/d\delta_k)^{-1}|_{\delta_k=0}=2d\Omega_m^2/\kappaex$
can be dramatically slowed by an amount that is tunable through
the optomechanical coupling strength $\Omega_m$. The quadratic and
cubic terms in $\keff$ characterize pulse absorption and group
velocity dispersion, respectively. In the relevant regime where
$\kappaex{\gg}\kappain,\kappaex{\gtrsim}\Omega_m$, these effects
are negligible within a bandwidth
$\Delta\omega{\sim}\textrm{min}\left(\frac{2\sqrt{2}\Omega_m^2}{\sqrt{N\kappaex\kappain}},\frac{2(6\pi)^{1/3}\Omega_m^2}{{\kappaex}N^{1/3}})\right)$.
The second term is the bandwidth over which certain frequency
components of the pulse acquire a $\pi$-phase shift relative to
others, leading to pulse distortion. This yields a bandwidth-delay
product of
\be
\Delta\omega\taud{\sim}\textrm{min}\left(\sqrt{2N\kappaex/\kappain},(6{\pi}N^2)^{1/3}\right)\label{eq:bwdelayproduct}
\ee
for static $\Omega_m$ and negligible mechanical losses. When intrinsic optical cavity
losses are negligible, and if one is
not concerned with pulse distortion, light can propagate over the full bandwidth ${\sim}4\Omega_m^2/\kappa$ of the slow-light polariton band and the bandwidth-delay product increases to $\Delta\omega\taud{\sim}N$~(see Appendix). On the other hand, we note that if
we had operated in a regime where $k_{0}d={\pi}n$, constructive
interference in reflection would limit the bandwidth-delay product
to $\Delta\omega\taud{\sim}1$, independent of system size.

In the static regime, the bandwidth-delay product obtained here is
analogous to CROW systems~\cite{scheuer05}. In the case of EIT, a
static bandwidth-delay product of
$\Delta\omega\tau_{\footnotesize\textrm{delay}}{\sim}\sqrt{\textrm{OD}}$
results, where OD is the optical depth of the atomic medium. This
product is limited by photon absorption and re-scattering into
other directions, and is analogous to our result
$\Delta\omega\taud{\sim}\sqrt{N\kappaex/\kappain}$ in the case of
large intrinsic cavity linewidth. On the other hand, when
$\kappain$ is negligible, photons are never lost and reflections
can be suppressed by interference. This yields an
improved scaling $\Delta\omega\taud{\sim}N^{2/3}$ or ${\sim}N$,
depending on whether one is concerned with group velocity
dispersion. In atomic media, the weak atom-photon coupling makes
achieving $\textrm{OD}>100$ very challenging~\cite{hong09}. In
contrast, in our system as few as $N{\sim}10$ elements would be
equivalently dense.

\subsection{Storage of optical pulse}

We now show that the group velocity
$v_{g}(t)=2d\Omega_m^2(t)/\kappaex$ can in fact be adiabatically
changed once a pulse is completely localized inside the system,
leading to distortion-less propagation at a dynamically tunable
speed. In particular, by tuning $v_{g}(t){\rightarrow}0$, the
pulse can be completely stopped and stored.

This phenomenon can be understood in terms of the
static band structure of the
system~(figure~\ref{fig:bandstructure}) and a ``dynamic
compression'' of the pulse bandwidth. The same
physics applies for CROW structures~\cite{yanik04,yanik05}, and
the argument is re-summarized here. First, under constant
$\Omega_m$, an optical pulse within the bandwidth of the polariton
band completely enters the medium. Once the pulse is inside, we
consider the effect of a gradual reduction in $\Omega_m(t)$.
Decomposing the pulse into Bloch wavevector components, it is
clear that each Bloch wavevector is conserved under arbitrary
changes of $\Omega_m$, as it is fixed by the system periodicity.
Furthermore, transitions to other bands are negligible provided
that the energy levels are varied adiabatically compared to the
size of the gap, which translates into an adiabatic condition
$|(d/dt)(\Omega_m^2/\kappa)|{\lesssim}\kappa^2$. Then,
conservation of the Bloch wavevector implies that the bandwidth of
the pulse is dynamically compressed, and the reduction in slope of
the polariton band~(figure~\ref{fig:bandstructure}) causes the pulse
to propagate at an instantaneous group velocity $v_{g}(t)$ without
any distortion. In the limit that $\Omega_m{\rightarrow}0$, the
polaritonic band becomes flat and completely mechanical in
character, indicating that the pulse has been reversibly and
coherently mapped onto stationary mechanical excitations within
the array. We note that since $\Omega_m$ is itself
set by the tuning cavities, its rate of change cannot exceed the
optical linewidth and thus the adiabaticity condition is always
satisfied in the weak-driving regime.

The maximum storage time is set by the mechanical decay rate,
${\sim}1/\gamma_m$. For realistic system parameters
$\omega_{m}/2\pi=10$~GHz and $Q_{m}=10^5$, this yields a storage
time of ${\sim}10\;\mu$s. In CROW structures,
light is stored as circulating fields in optical nano-cavities,
where state of the art quality factors of $Q{\sim}10^6$ limit the
storage time to ${\sim}1$~ns. The key feature of our system is
that we effectively ``down-convert'' the high-frequency optical
fields to low-frequency mechanical excitations, which naturally
decay over much longer time scales. While storage
times of ${\sim}10$~ms are possible using atomic
media~\cite{figueroa06}, their bandwidths so far have been limited
to $<1$~MHz~\cite{yanik05}. In our system, bandwidths of
${\sim}1$~GHz are possible for realistic circulating powers in the
tuning cavities.

\subsection{Imperfections in storage}

The major source of error in our device will be mechanical noise,
which through the optomechanical coupling can be mapped into noise
power in the optical waveguide output. In our system, mechanical
noise emerges via thermal fluctuations and Stokes
scattering~(corresponding to the counter-rotating terms in the
optomechanical interaction that we omitted from
Equation~(\ref{eq:Hlinear})). To analyze these effects, it suffices to
consider the case of static~$\Omega_m$, and given the linearity of
the system, no waveguide input~(such that the output will be
purely noise). For a single array element, the optomechanical
driving $\Omega_m$ results in optical cooling of the mechanical
motion~\cite{wilson-rae07,marquardt07}, with the mechanical energy
$E_{m}$ evolving as~(see Appendix)
\be
\frac{dE_m}{dt}=-\gamma_m\left(E_m-\hbar\omega_{m}\bar{n}_{\footnotesize\textrm{th}}\right)-{\Gammaopt}E_{m}+\Gammaopt\frac{\kappa^2}{\kappa^2+16\omega_m^2}\left(E_m+\hbar\omega_m\right).\label{eq:dEmdt}
\ee
The first term on the right describes equilibration with the thermal surroundings, where
$\bar{n}_{\footnotesize\textrm{th}}=(e^{\hbar\omega_{m}/k_{B}T_{b}}-1)^{-1}$ is the Bose occupation number at the
mechanical frequency and $T_b$ is the bath temperature. The second~(third) term corresponds to cooling~(heating) through
anti-Stokes~(Stokes) scattering, with a rate proportional to $\Gammaopt=4\Omega_m^2/\kappa$. The Stokes process is
suppressed relative to the anti-Stokes in the limit of good sideband resolution $\kappa/\omega_m{\ll}1$. For an array of
$N$ elements, a simple upper bound for the output noise power at one end of the waveguide is given by
$\Pnoise=(1/2)({\Gammaopt}E_{ss})N(\omega_1/\omega_m)(\kappaex/\kappa)$, where $E_{ss}$ is the steady-state solution of
Equation~(\ref{eq:dEmdt}). The factor of $1/2$ accounts for the optical noise exiting equally from both output directions,
${\Gammaopt}E_{ss}$ is the optically-induced mechanical energy dissipation rate, and $\kappaex/\kappa$ describes the
waveguide coupling efficiency. The term $\omega_1/\omega_m$ represents the transduction of mechanical to optical energy
and is essentially the price that one pays for down-converting optical excitations to mechanical to yield longer storage
times -- in turn, any mechanical noise gets ``up-converted'' to optical energy~(whereas the probability of having a
thermal optical photon is negligible).  In the relevant regime where $\Gammaopt{\gg}\gamma_m$,
\be
\Pnoise{\approx}\frac{N\hbar\omega_1}{2}\frac{\kappaex}{\kappa}\left(\gamma_{m}\bar{n}_{\footnotesize\textrm{th}}+\Gammaopt\left(\frac{\kappa}{4\omega_m}\right)^2\right).
\ee
This noise analysis is valid only in the weak-driving
regime~($\Omega_m{\lesssim}\kappa$)~\cite{wilson-rae07,marquardt07}.
The strong driving regime, where the mechanical motion acquires
non-thermal character~\cite{genes08} and can become entangled with
the optical fields~\cite{genes08c}, will be treated in future
work.

At room temperature,
$\bar{n}_{\footnotesize\textrm{th}}{\approx}k_{B}T_b/\hbar\omega_m$
is large and thermal noise will dominate, yielding a noise power
of ${\sim}0.4$~nW per element for previously given system
parameters and $\kappaex/\kappa{\approx}1$. This is independent of
$\Gammaopt$ provided that $\Gammaopt{\gg}\gamma_m$, which reflects
the fact that all of the thermal heating is removed through the
optical channel.For high temperatures, the
thermal noise scales inversely with $\omega_m$, and the use of
high-frequency mechanical oscillators ensures that the noise
remains easily tolerable even at room temperature.

Thermal noise in the high-frequency oscillator can essentially be
eliminated in cryogenic environments, which then enables faithful
storage of single photons. Intuitively, a
single-photon pulse can be stored for a period only as long as the
mechanical decay time $\sim\gamma_{m}^{-1}$, and as long as a
noise-induced mechanical excitation is unlikely to be generated
over a region covering the pulse length and over the transit time
$\taud$. The latter condition is equivalent to the statement that
the power
$P_{\footnotesize\textrm{ph}}{\sim}\hbar\omega_1\Delta\omega$ in
the single-photon pulse exceeds $\Pnoise$. While we have focused
on the static regime thus far, when thermal heating is negligible,
realizing $P_{\footnotesize\textrm{ph}}/\Pnoise{\gtrsim}1$ in the
static case in fact ensures that the inequality holds even when
$\Gammaopt(t)$ is time-varying. Physically, the rate of Stokes
scattering scales linearly with $\Gammaopt$ while the group
velocity scales inversely, and thus the probability of a noise
excitation being added on top of the single-photon pulse is fixed
over a given transit length.

In a realistic setting, the optomechanical driving amplitude
$\Omega_m$ itself will be coupled to the bath temperature, as
absorption of the pump photons in the tuning cavities leads to
material heating. To understand the limitations as a quantum
memory, we have numerically optimized the static bandwidth-delay
product $\Delta\omega\taud$ for a train of single-photon pulses,
subject to the constraints
$\Delta\omega<\textrm{min}\left(2\sqrt{2}\Omega_m^2/\sqrt{N\kappaex\kappain},2(6\pi)^{1/3}\Omega_m^2/({\kappaex}N^{1/3})\right)$,
$P_{\footnotesize\textrm{ph}}/\Pnoise>1$, and $\gamma_{m}\taud<1$.
As a realistic model for the bath temperature, we take
$T_b=T_0+\chi\alpha_2^2=T_0+\chi(\Omega_m/h)^2$, where $T_0$ is
the base temperature and $\chi{\sim}2\;\mu$K is a temperature
coefficient that describes heating due to pump absorption~(see
Appendix). Using $T_0=100$~mK and $Q_m=10^5$, we find
$(\Delta\omega\taud)_{\footnotesize\textrm{max}}{\sim}110$, which
is achieved for parameter values $N{\sim}275$,
$\kappaex/2\pi{\sim}$1.1~GHz, and $\Omega_m/2\pi{\sim}130$~MHz.

\section{Optomechanical crystal design}\label{sec:design}
\begin{figure}[htbp]
\begin{center}
\includegraphics[width=17cm]{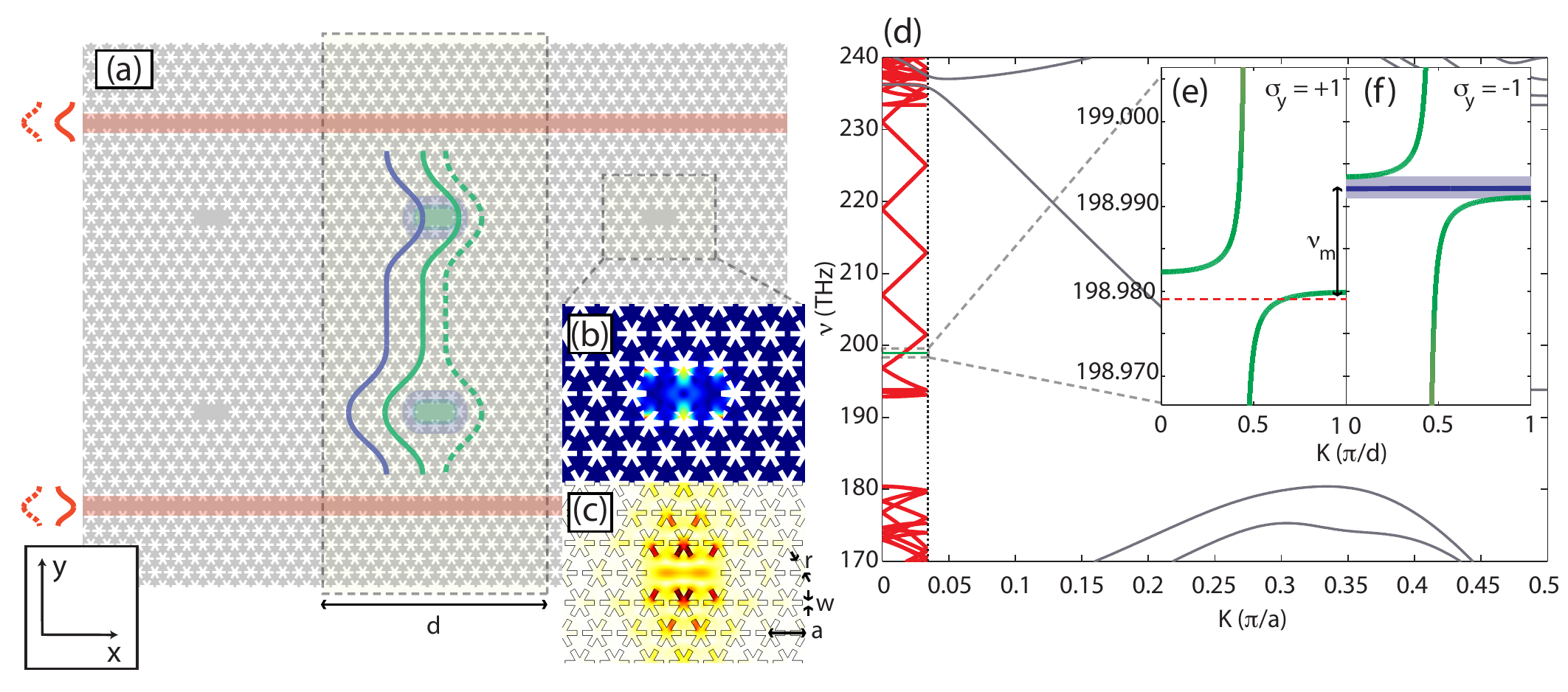}
\end{center}
\caption{(a) Top view of the proposed optomechanical crystal array, with the superlattice unit-cell of length $d$
  highlighted in the center. The unit cell contains two coupled L2 defect cavities (shaded in grey) with two
  side-coupled linear defect optical waveguides (shaded in red).  The different envelope functions pertain to the odd and
  even optical cavity supermodes (green solid and dashed lines, respectively), the odd mechanical supermode (blue solid
  line), and the odd and even optical waveguide modes (red solid and dashed lines).  The displacement field amplitude
  $|\mathbf{Q}(\mathbf{r})|$ of the mechanical mode and in-plane electric field amplitude $|\mathbf{E}(\mathbf{r})|$ of
  the optical mode are shown in (b) and (c), respectively, for a single L2 defect cavity. (d) Bandstructure of the
  linear-defect waveguide (grey) and the zone folded superlattice of the entire coupled-resonator system (red). The
  cavity mode (green) crosses the superlattice band at mid-zone, and the waveguide-cavity interaction is shown in more
  detail in the insets (e) and (a) for the even ($\sigma_y = +1$) and odd ($\sigma_y = -1$) supermodes, respectively.
 \label{fig:OMC_model}}
\end{figure}

A schematic showing a few periods of our proposed 2D OMC slow-light structure is given in figure~\ref{fig:OMC_model}.  The structure is built around a ``snowflake'' crystal pattern of etched holes into a Silicon slab~\cite{safavi-naeini10a}.  This pattern, when implemented with a physical lattice constant of $a=400~\text{nm}$, snowflake radius $r = 168~\text{nm}$, and snowflake width $w = 60~\text{nm}$~(see figure~\ref{fig:OMC_model}(a)), provides a simultaneous
phononic bandgap from $8.6$ to $12.6$ GHz and a photonic pseudo-bandgap from $180$ to $230$ THz~(see Appendix). Owing to its unique bandgap properties, the snowflake patterning can be used to form waveguides and resonant cavities for both acoustic and optical waves simply by removing regions of the pattern.  For instance, a single point defect, formed by removing two adjacent holes~(a so-called ``L2'' defect), yields the co-localized phononic and photonic resonances shown in figures~\ref{fig:OMC_model}(b) and (c), respectively.  The radiation pressure, or optomechanical coupling between the two resonances can be quantified by a coupling rate, $g$, which corresponds to the frequency shift in the optical resonance line introduced by a single phonon of the mechanical resonance. Numerical finite-element-method (FEM) simulations of the L2 defect indicate the mechanical resonance occurs at $\omega_m/2\pi = 11.2~\text{GHz}$, with a coupling rate of $g/2\pi = 489~\text{kHz}$ to the optical mode at frequency $\omega_o/2\pi = 199~\text{THz}$ (free-space optical wavelength of $\lambda_0 \approx 1500~\text{nm}$).

In order to form the double-cavity system described in the slow-light scheme above, a pair of L2 cavities are placed in the near-field of each other as shown in the dashed box region of figure~\ref{fig:OMC_model}(a).  Modes of the two degenerate L2 cavities mix, forming supermodes of the double-cavity system which are split in frequency.  The frequency splitting between modes can be tuned via the number of snowflake periods between the cavities.  As described in more detail in the Appendix, it is the optomechanical cross-coupling of the odd ($\m E_-$) and even ($\m E_+$) optical supermodes mediated by the motion of the odd parity mechanical supermode ($\m Q_-$) of the double-cavity that drives the slow-light behaviour of the system.  Since $\m Q_-$ is a displacement field that is antisymmetric about the two cavities, there is no optomechanical self-coupling between the optical supermodes and this mechanical mode. On the other hand, the cross-coupling between the two different parity optical supermodes is large and given by $h=g/\sqrt{2}=2\pi(346~\text{kHz})$. By letting $\op{a}{1}$, $\op{a}{2}$ and $\op{b}{}$ be the annihilation operators for the modes $\m E_-$, $\m E_+$ and $\m Q_-$, we obtain the system Hamiltonian of equation~(\ref{eq:H}).

The different spatial symmetries of the optical cavity supermodes also allow them to be addressed independently. To
achieve this we create a pair of linear defects in the snowflake lattice as shown in figure~\ref{fig:OMC_model}(a), each
acting as a single-mode optical waveguide at the desired frequency of roughly $200~\text{THz}$ (see
figure~\ref{fig:OMC_model}(d). Sending light down both waveguides, with the individual waveguide modes either in or out of
phase with each other, will then excite the even or odd supermode of the double cavity, respectively.  The waveguide
width and proximity to the L2 cavities can be used to tune the cavity loading~(see Appendix), which for the structure in
figure~\ref{fig:OMC_model}(a) results in the desired $\kappaex/2\pi = 2.4~\text{GHz}$. It should be noted that these
line-defect waveguides do not guide phonons at the frequency of $\m Q_-$, and thus no additional phonon leakage
is induced in the localized mechanical resonance.

The full slow-light waveguide system consists of a periodic array of the double-cavity, double-waveguide structure.  The
numerically computed band diagram, for spacing $d=15a$ periods of the snowflake lattice between cavity elements (the
superlattice period), is shown in figure~\ref{fig:OMC_model}(d).  This choice of superlattice period results in the folded
superlattice band intersecting the $\m E_-$~($\op{a}{1}$) cavity frequency $\omega_1$ at roughly mid-zone, corresponding
to the desired inter-cavity phase shift of $kd = \pi/2$.  A zoom-in of the bandstructure near the optical cavity
resonances is shown in figures~\ref{fig:OMC_model}(e) and (f).  In figure~\ref{fig:OMC_model}(e) the even parity supermode
bandstructure is plotted~(i.e., assuming the even supermode of the double-waveguide is excited), whereas in
figure~\ref{fig:OMC_model}(f) it is the odd parity supermode bandstructure. 
\begin{figure}[htbp]
\begin{center}
\includegraphics[width=17cm]{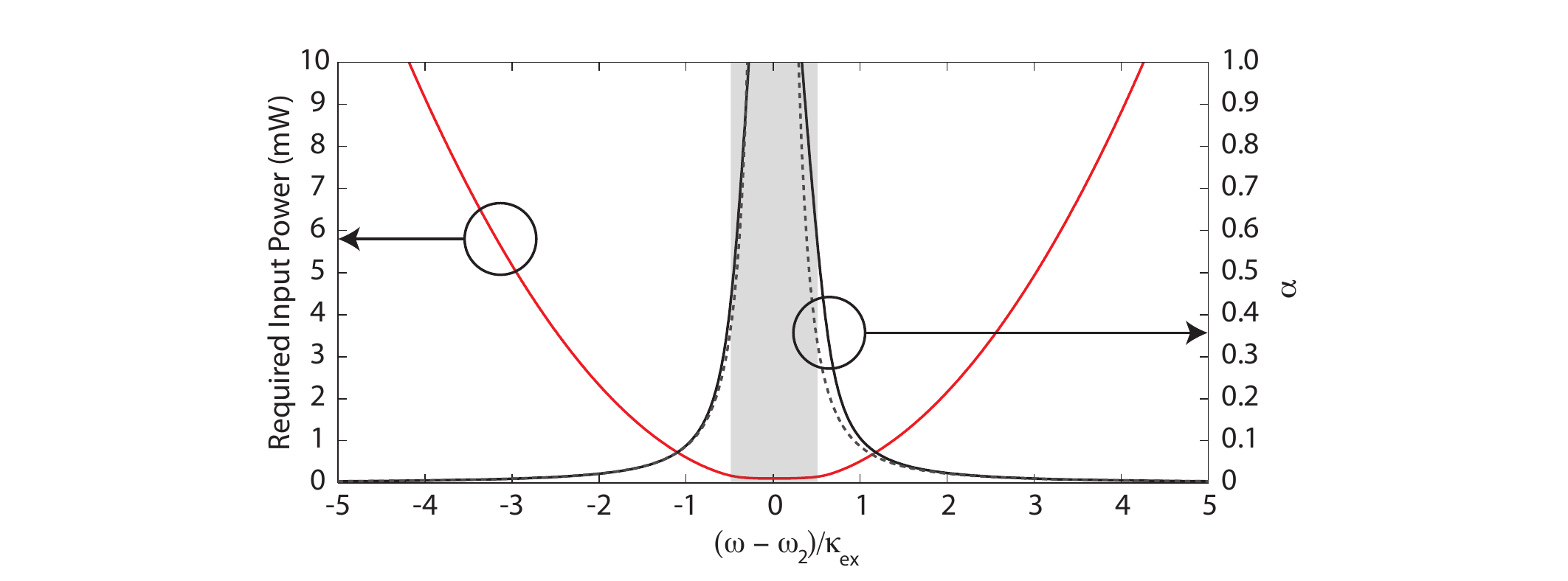}
\end{center}
\caption{The input power (red line) required to achieve the system
parameters used in the text, i.e. $\Omega_m/2\pi=130~\text{MHz}$
with $h/2\pi = 0.346~\text{MHz}$, and the attenuation per unit cell
$\alpha$ (solid black line) are shown as a function of detuning of
the pump beam from the pump cavity frequency. The dotted line is
the approximate expression derived for the attenuation, $\alpha
\approx \kappaex \kappain / 4\delta_k^2$. The grey region indicates the band gap in which the pump cavities cannot be excited from the waveguide. The trade-off between
small pump input powers and low pump attenuation factors  is
readily apparent in this plot.
 \label{fig:inputpower}}
\end{figure}

 A subtlety in the optical pumping of the
periodically arrayed waveguide system is that for the $\m E_+$~($\op{a}{2}$) optical cavity resonance at $\omega_2$,
there exists a transmission bandgap.  In order to populate cavity $\op{a}{2}$, then, and to create the polaritonic band at
$\omega_1$, the pump beam must be slightly off-resonant from $\omega_2$, but still at $\omega_1 - \omega_m$. We achieve
this by choosing a double-cavity separation ($14$ periods) resulting in a cavity mode splitting~($(\omega_1 -
\omega_2)/2\pi = 9.7~\text{GHz}$) slightly smaller than the mechanical frequency~($\omega_m/2\pi = 11.2~\text{GHz}$), as
shown in figures~\ref{fig:OMC_model}(e) and (f).  By changing the detuning between $\omega_1-\omega_2$ and $\omega_m$, a trade-off can be made between the attenuation of the pump beam per unit cell, $\alpha \equiv \exp(-\text{Im}\{K\} d)$, and total required input power, shown in figure~\ref{fig:inputpower}. 
In \ref{app:input} we show that the total attenuation per unit cell, is given by  $\alpha
\approx \kappaex \kappain / 4\delta_k^2$. Interestingly, by using higher input powers such that $\alpha \rightarrow 0$, it is in principle possible to eliminate completely effects due to absorption in the array, which may lead to inhomogeneous pump photon occupations.

\section{Outlook}

The possibility of using optomechanical systems to facilitate major tasks in classical optical networks has been
suggested in several recent proposals~\cite{rosenberg09,lin09,safavi-naeini10a}. This present work not only extends
these prospects, but proposes a fundamentally new direction where optomechanical systems can be used to control and
manipulate light at a quantum mechanical level. Such efforts would closely mirror the many proposals to perform similar
tasks using EIT and atomic ensembles~\cite{fleischhauer05}.  At the same time, the optomechanical array has a number of
novel features compared to atoms, in that each element can be deterministically positioned, addressed, and manipulated,
and a single element is already optically dense. Furthermore, the ability to freely convert between phonons and photons enables new possibilities for manipulating light through the manipulation of sound. Taken together, this raises the possibility that mechanical systems can
provide a novel, highly configurable on-chip platform for realizing quantum optics and ``atomic'' physics.

\section{Acknowledgements}
This work was supported by the DARPA/MTO ORCHID program through a grant from AFOSR.  DC acknowledges support from the NSF and the Gordon and Betty Moore Foundation through Caltech's Center for the Physics of Information. ASN acknowledges support from NSERC. M.H. acknowledges support from the U.S. Army Research Office MURI award W911NF0910406. 

\appendix

\section{Equations of motion for optomechanical crystal array}

Here we derive the equations of motion for an array of
optomechanical systems coupled to a two-way waveguide.  Because
each element in the array couples independently to the waveguide,
it suffices here to only consider a single element, from which the
result for an arbitrary number of elements is easily generalized.

We model the interaction between the active cavity mode $1$ and
the waveguide with the following Hamiltonian,
\bea
H_{\footnotesize\textrm{cav-wg}}&=&\int_{-\infty}^{\infty}\,dk\,\hbar(ck-\omega_1)\opdagger{a}{R,k}\op{a}{R,k}-\int_{-\infty}^{\infty}\,dk\,\hbar(ck+\omega_1)\opdagger{a}{L,k}\op{a}{L,k}-\nonumber\\&&
{\hbar}g\sqrt{2\pi}\int_{-\infty}^{\infty}\,dz\,\delta(z-z_j)\left(\opdagger{a}{1}\left(\op{a}{R}(z)+\op{a}{L}(z)\right)+h.c.\right).\label{eqSI:Hcavwg}
\eea
Here $\op{a}{R,k},\op{a}{L,k}$ are annihilation operators for
left- and right-going waveguide modes of wavevector $k$, and $\omega_1$ is the frequency of the cavity mode. For convenience we have defined all
optical energies relative to $\omega_1$, and assumed that the waveguide has a linear dispersion relation $\omega(k)=c|k|$. The last term on the right describes a point-like coupling between the cavity~(at position $z_j$) and the left- and right-going waveguide modes, with a strength $g$. The operator $\op{a}{R}(z)$ physically describes the annihilation of a
right-going photon at position $z$ and is related to the wavevector annihilation operators by $\op{a}{R}(z)=\frac{1}{\sqrt{2\pi}}\int_{-\infty}^{\infty}\,dk\,e^{ikz}\op{a}{R,k}$~(with a similar definition for $\op{a}{L}(z)$). Eq.~(\ref{eqSI:Hcavwg}) resembles a standard Hamiltonian used to formulate quantum cavity input-output relations~\cite{gardiner85}, properly generalized to the case when the cavity accepts an input from either direction. Note that we make the approximation that the left- and right-going waves can be treated as separate quantum fields, with modes in each direction running from $-\infty<k<\infty$. This allows both the left- and right-going fields to separately satisfy canonical field commutation relations, $\left[\op{a}{R}(z),\op{a}{R}(z')\right]=\left[\op{a}{L}(z),\op{a}{L}(z')\right]=\delta(z-z')$, while commuting with each other. Thus each field contains some unphysical modes~(\textit{e.g.}, wavevector components $k<0$ for the right-going field), but the approximation remains valid as long as there is no process in the system evolution that allows for the population of such modes.

From the Hamiltonian above, one finds the following Heisenberg
equation of motion for the right-going field,
\be
\left(\frac{1}{c}\frac{\partial}{{\partial}t}+\frac{\partial}{{\partial}z}\right)\op{a}{R}(z)=\frac{\sqrt{2\pi}ig}{c}\delta(z-z_j)\op{a}{1}+ik_{0}\op{a}{R},\label{eqSI:waveeqn}
\ee
where $k_0=\omega_1/c$. A similar equation holds for $\op{a}{L}$. The coupling of the cavity mode to a
continuum of waveguide modes leads to irreversible decay of the
cavity at a rate $\kappaex$. Below, we will show that
$\kappaex$ is related to the parameters in the Hamiltonian by
$\kappaex=4{\pi}g^2/c$. With this identification, one recovers
Eq.~(5) in the main text.

The Heisenberg equation of motion for the cavity mode is given by
\be
\frac{d}{dt}\op{a}{1}=ig\sqrt{2\pi}\left(\op{a}{R}(z_j)+\op{a}{L}(z_j)\right).
\ee
To cast this equation into a more useful form, we first integrate the field
equation~(\ref{eqSI:waveeqn}) across the discontinuity at $z_j$,
\bea \op{a}{R}(z_j^{+}) & = &
\op{a}{R}(z_j^{-})+\frac{\sqrt{2\pi}ig}{c}\op{a}{1}, \\
\op{a}{L}(z_j^{-}) & = &
\op{a}{L}(z_j^{+})+\frac{\sqrt{2\pi}ig}{c}\op{a}{1}. \eea
We can define $\op{a}{L,in}(z_j)=\op{a}{L}(z_j^{+})$ and
$\op{a}{R,in}(z_j)=\op{a}{R}(z_j^{-})$ as the input fields to the
cavity. It then follows that
\be
\frac{d}{dt}\op{a}{1}=ig\sqrt{2\pi}\left(\op{a}{R,in}+\op{a}{L,in}\right)-\frac{2{\pi}g^2}{c}\op{a}{1},
\ee
and thus we indeed see that the waveguide induces a cavity decay rate
$\kappaex/2=2{\pi}g^2/c$. In the case where the cavity has an
additional intrinsic decay rate $\kappain$, a similar derivation
holds to connect the intrinsic decay with some corresponding noise input field
$\op{a}{in}$. From these considerations, and including the
opto-mechanical coupling, one arrives at Eq.~(3) in the main text,
\be
\frac{d\hat{a}_{1}}{dt}=-\frac{\kappa}{2}\hat{a}+i\Omega_{m}\hat{b}+i\sqrt{\frac{c\kappaex}{2}}\left(\op{a}{R,in}(z_j)+\op{a}{L,in}(z_j)\right)+\sqrt{c\kappain}\op{a}{N}(z_j).\label{eqSI:dadt}
\ee

Finally, we consider the equation of motion for the mechanical
mode given by Eq.~(4) in the main text,
\be
\frac{d\hat{b}}{dt}=-\frac{\gamma_m}{2}\hat{b}+i\Omega_{m}\hat{a}_{1}+\op{F}{N}(t).\label{eqSI:dbdt}
\ee
The zero-mean noise operator $\op{F}{N}$ must accompany the decay term in the mechanical evolution in order to preserve canonical
commutation relations of $\hat{b},\hat{b}^{\dagger}$ at all times. In the case where the decay causes the
mechanical motion to return to thermal equilibrium with some
reservoir at temperature $T_b$, the noise operator has a two-time
correlation function given by
$\avg{F_{N}(t)F_{N}^{\dagger}(t')}=\gamma_{m}(\bar{n}+1)\delta(t-t')$~\cite{meystre99},
where $\bar{n}=\left(e^{\hbar\omega_{m}/k_{B}T_b}-1\right)^{-1}$
is the Bose occupation number at the mechanical frequency
$\omega_m$.

\section{Transfer matrix analysis of propagation}

First we derive the reflection and transmission coefficients for a
single element in the case of constant opto-mechanical driving
amplitude $\Omega_m$. Given the linearity of the system, it
suffices to treat Eqs.~(\ref{eqSI:waveeqn}),~(\ref{eqSI:dadt}),
and~(\ref{eqSI:dbdt}) as classical equations for this purpose, and furthermore to set the noise terms
$\op{F}{N}=0$, $\op{a}{in}=0$. For concreteness, we will consider
the case of an incident right-going cw field in the waveguide.
Upon interaction with the opto-mechanical system at $z_j$, the
total right-going field can be written in the form
\be
a_{R}(z)=e^{ikz-i\delta_{k}t}\left(\Theta(-z+z_j)+t(\delta_k)\Theta(z-z_j)\right),
\ee
while the left-going field is given by
$a_{L}(z)=e^{-ikz-i\delta_{k}t}r(\delta_k)\Theta(-z+z_j)$. Here
$\Theta(z)$ is the unit step function, $\delta_k=ck-\omega_1$ is
the detuning of the input field from the cavity resonance, and
$r,t$ are the reflection and transmission coefficients for the
system. At the same time, we look for solutions of the cavity
field and mechanical mode of the form
$a_{1}=Ae^{-i\delta_{k}t}$ and
$b=Be^{-i\delta_{k}t}$. The coefficients $r,t,A,B$
can be obtained by substituting this ansatz into
Eqs.~(\ref{eqSI:waveeqn}),~(\ref{eqSI:dadt}), and~(\ref{eqSI:dbdt}).
This yields the reflection coefficient given by Eq.~(6) in the
main text, while the transmission coefficient is related by
$t=1+r$.

To calculate propagation through an array of $N$ elements, it is
convenient to introduce a transfer matrix formalism. Specifically,
the fields immediately to the right of the opto-mechanical
element~(at $z=z_j^{+}$) can be related to those immediately to
the left~(at $z=z_j^{-}$) in terms of a transfer matrix $M_{om}$,
\be \left(\begin{array}{c} a_{R}(z_j^{+}) \\ a_{L}(z_j^{+})
\end{array}\right)=M_{om}\left(\begin{array}{c} a_{R}(z_j^{-}) \\ a_{L}(z_j^{-})
\end{array}\right), \ee
where
\be M_{om}=\frac{1}{t}\left(\begin{array}{cc} t^2-r^2 & r \\ -r &
1 \end{array}\right). \label{eqSI:Mom} \ee
On the other hand, free propagation in the waveguide is
characterized by the matrix $M_f$,
\be \left(\begin{array}{c} a_{R}(z+d) \\ a_{L}(z+d)
\end{array}\right)=M_{f}\left(\begin{array}{c} a_{R}(z) \\ a_{L}(z)
\end{array}\right), \ee
where
\be M_{f}=\left(\begin{array}{cc} e^{ikd} & 0 \\ 0 & e^{-ikd}
\end{array}\right).\label{eqSI:Mf}  \ee
The transfer matrix for an entire system can then be obtained by
successively multiplying the transfer matrices for a single
element and for free propagation together. In particular, the
transfer matrix for a single ``block'', defined as interaction
with a single opto-mechanical element followed by free propagation
over a distance $d$ to the next opto-mechanical element, is given
by $M_{block}=M_{f}M_{om}$, and the propagation over $N$ blocks is
simply characterized by $M_N=M_{block}^N$.

Before studying the propagation through the entire array, we first
focus on the propagation past two blocks, $M_2=M_{block}^2$.
Because we want our device to be highly transmitting when the optomechanical coupling is turned on, we choose
the spacing $d$ between consecutive blocks to be such that
$k_{0}d=\frac{\pi}{2}(2n+1)$, where $n$ is an integer. Physically,
this spaces consecutive elements by an odd multiple of
$\lambda/4$, where $\lambda$ is the resonant wavelength, such that
the reflections from consecutive elements tend to destructively
interfere with each other. This can be confirmed by examining the
resulting reflection coefficient for the two-block system,
\be
r_{2}\equiv\frac{M_{2}(1,2)}{M_{2}(2,2)}=-\frac{\kappaex^2\delta_k^2}{2\Omega_m^4}+O(\delta_k^3),
\ee
where $M_{2}(i,j)$ denotes matrix elements of $M_2$. Note that the
reflection coefficient is now suppressed as a quadratic in the
detuning, whereas for a single element
$r{\approx}i\kappaex\delta_k/2\Omega_m^2$ is linear. In the
above equation, we have made the simplifying approximation that
$kd{\approx}k_{0}d$, since in realistic systems the dispersion from free propagation will be negligible compared to that arising from interaction with an opto-mechanical element.

Now we can consider transmission past $N/2$ pairs of two
blocks~(\textit{i.e.}, $N$ elements in total). Because the
reflection $r_2$ is quadratic in the detuning, its effect on the
total transmission is only of order $\delta_k^4$~(because the
lowest order contribution is an event where the field is reflected
twice before passing through the system). Thus, up to
$O(\delta_k^3)$, the total transmission coefficient $t_{N}$ is
just given by $t_{N}{\approx}t_{2}^{N/2}$, where
$t_{2}=1/M_{2}(2,2)$ is the transmission coefficient for a
two-block system. It is convenient to write $t_{N}=e^{i{\keff}Nd}$
in terms of an effective wavevector $\keff$, which leads to
Eq.~(7) in the main text. Performing a similar analysis for the
case where $k_{0}d=n\pi$, where reflections from consecutive
elements interfere constructively, one finds that the
bandwidth-delay product for the system $\Delta\omega\taud{\sim}1$
does not improve with the system size.

\section{Optical noise power}

In this section we derive the optical cooling equation given by
Eq.~(9) in the main text. We begin by considering the system
Hamiltonian for a single element~(Eq.~(1) of main text), in the case where the tuning mode $2$ is driven on resonance and can be approximated by a classical field, $\op{a}{2}{\rightarrow}\alpha_{2}e^{-i\omega_{2}t}$,
\be \tilde{H}_{om}=\hbar\omega_{1}\opdagger{a}{1}\op{a}{1}+\hbar\omega_{m}\hat{b}^{\dagger}\hat{b}+{\hbar}h\left(\hat{b}+\hat{b}^{\dagger}\right)\left(\opdagger{a}{1}\alpha_{2}e^{-i\omega_{2}t}+\alpha_{2}^{\ast}e^{i\omega_{2}t}\op{a}{1}\right).\label{eqSI:H} \ee
Defining a detuning $\delta_L=\omega_2-\omega_1$ indicating the frequency difference between the two cavity modes, we can re-write Eq.~(\ref{eqSI:H}) in a rotating frame,
\be \tilde{H}_{om}=-\hbar\delta_L\opdagger{a}{1}\op{a}{1}+\hbar\omega_{m}\hat{b}^{\dagger}\hat{b}+{\hbar}\Omega_{m}\left(\hat{b}+\hat{b}^{\dagger}\right)\left(\op{a}{1}+\opdagger{a}{1}\right),\label{eqSI:Hrot} \ee
where $\Omega_m=h\alpha_2$~(we have re-defined the phases such that $\Omega_m$ is real). In the weak driving limit~($\Omega_m{\lesssim}\kappa$), the cavity dynamics can be formally eliminated to arrive at effective optically-induced cooling equations for the mechanical motion~\cite{wilson-rae07,marquardt07}. In particular, the opto-mechanical coupling terms $\opdagger{a}{1}\hat{b}+h.c.$ and $\opdagger{a}{1}\hat{b}^{\dagger}+h.c.$ induce anti-Stokes and Stokes scattering, respectively. These processes yield respective optically-induced  cooling~($\Gamma_{-}$) and heating~($\Gamma_{+}$) rates
\be \Gamma_{\mp}=\frac{\kappa\Omega_m^2}{(\delta_L{\pm}\omega_m)^2+(\kappa/2)^2}. \ee
In the case where $\delta_L=-\omega_m$, the cooling process is resonantly enhanced by the cavity, yielding a cooling rate $\Gammaopt\equiv\Gamma_{-}(\delta_L=-\omega_m)=4\Omega_m^2/\kappa$ as given in the main text. Also in this case, the optical heating rate is given by $\Gamma_{+}=\Gammaopt\frac{\kappa^2}{\kappa^2+16\omega_m^2}$. This leads to the net cooling dynamics given by Eq.~(9) in the main text,
\be
\frac{dE_m}{dt}=-\gamma_m\left(E_m-\hbar\omega_{m}\bar{n}_{th}\right)-{\Gammaopt}E_{m}+\Gammaopt\frac{\kappa^2}{\kappa^2+16\omega_m^2}\left(E_m+\hbar\omega_m\right).\label{eqSI:dEmdt}
\ee
Because the optical cooling process removes phonons from the mechanical system via optical photons that leak out of the cavity, one can identify $(\omega_1/\omega_m){\Gammaopt}E_m$ as the amount of optical power that is being leaked by the cavity in the anti-Stokes sideband during the cooling process. Similarly, the cavity leaks an amount of power $(\omega_1/\omega_m){\Gammaopt}\frac{\kappa^2}{\kappa^2+16\omega_m^2}(E_m+\hbar\omega_m)$ in the Stokes sideband. We have ignored this contribution in Eq.~(10) in the main text, because its large frequency separation~($2\omega_m$) from the signal allows it to be filtered out, but otherwise it approximately contributes an extra factor of $2$ to the last term in Eq.~(10). Finally, we remark that the expression for $\Pnoise$ given by Eq.~(10) represents an upper bound in that it does not account for the possibility that the output spectrum from a single element may exceed the transparency bandwidth, which could cause some
light to be absorbed within the system after multiple reflections and not make it to the end of the waveguide.

\section{Band structure analysis}

\subsection{Derivation of dispersion relation}

For simplicity, here we work only with the classical equations so that the intrinsic noise terms in the Heisenberg-Langevin equations can be ignored. We begin by transforming Eqs.~(\ref{eqSI:dadt}) and (\ref{eqSI:dbdt}) to the Fourier domain,
\bea 0 &=&\left(i\delta_k-\frac{\kappa}{2}\right)a +i\Omega_{m}b+i\sqrt{\frac{c\kappaex}{2}}\left(a_{R,in}(z_j)+a_{L,in}(z_j)\right),\label{eqSI:dadt0}\\
 0 & = & \left(i\delta_k-\frac{\gamma_m}{2}\right)b+i\Omega_{m}a.
\eea
To simplify the notation, we define operators at the boundaries of the unit cells~(immediately to the left of an optomechanical element) given by ${c}_{j} = - i \sqrt{c} {a}_{R}(z_j^{-})$, ${d}_{j} = - i \sqrt{c} {a}_{L}(z_j^{-})$. It is also convenient to re-write the transfer matrix $M_{om}$ in the form
\be
M_{om}  =  \left(\begin{array}{cc} 1-\beta & -\beta \\ \beta & 1+\beta \end{array}\right), \label{eqSI:newMom}
\ee
with the parameter $\beta(\delta_k)$ given by
\be
\beta(\delta_k) = \frac{-i\kappaex\delta_k}{-i\kappain\delta_k + 2(\Omega_m^2 -\delta_k^2)}\label{eqSI:beta}.  \ee

The transfer matrix $M_{block}$ describing propagation to the next unit cell can subsequently be diagonalized, $M_{block} = S D S^{-1}$, with the diagonal matrix $D$ given by
\be D=\left(\begin{array}{cc} e^{iKd} & 0 \\ 0 & e^{-iKd} \end{array}\right).\label{eqSI:D} \ee
Physically, this diagonalization corresponds to finding the Bloch wavevectors $K(\delta_k)$ of the periodic system. The dispersion relation for the system can be readily obtained through the equation
\be
\cos (K(\delta_k) d) = \cos(k d) - i \beta(\delta_k) \sin (k d)\label{eqSI:dispersion}.
\ee
Writing $k$ in terms of $\delta_k$, we arrive at $kd = \omega_1 d / c + \delta_k d /c$. As described previously, the desirable operation regime of the system is such that the phase imparted in free propagation should be  $\omega_{1}d/c=(2n+1)\pi/2$. For concreteness, we set here $\omega_1 d / c = \pi/2$, satisfying this condition. For the frequencies $\delta_k$ of interest, which easily satisfy the condition $|\delta_k| \ll d/c$ and ignoring the intrinsic loss $\kappain$, the simple approximate dispersion formula
\be
\cos (K(\delta_k) d) = - \frac{\kappaex \delta_k}{2(\Omega_m^2 - \delta_k^2)}
\ee
can be found. This dispersion relation yields two bandgaps, which extend from $\pm \kappaex/2$ and $\pm 2 \Omega_m^2 / \kappa$, in the weakly coupled EIT regime~$(\Omega_m \lesssim \kappaex)$. We therefore have three branches in the band structure, with the narrow central branch having a width of $4\Omega_m^2 / \kappaex$. This branch has an optically tunable width and yields the slow-light propagation.

The dispersive and lossy properties of the array can also be found by analyzing Eq.~(\ref{eqSI:dispersion}) perturbatively. Expanding Eq.~(\ref{eqSI:dispersion}) as a power series in $\delta_k$, we find
\be K(\delta_k) = k_0+\frac{\kappaex\delta_k}{2d\Omega_m^2}+\frac{i\kappaex\kappain\delta_k^2}{4d\Omega_m^4}+\frac{(2\kappaex^3-3\kappaex\kappain^2+12\kappaex\Omega_m^2)\delta_k^3}{24d\Omega_m^6} + O(\delta_k^4),\label{eqSI:K} \ee
which agrees with Eq.~(7) in the main text.

\subsection{Fractional Occupation Calculation}

In our system, the Bloch functions are hybrid waves arising from
the mixing of optical waveguide, optical cavity and mechanical
cavity excitations. It is therefore of interest to calculate the
hybrid or polaritonic properties of these waves, by studying the
energy distribution of each Bloch mode.

The number of photons $n_\text{WG}$ in the waveguide can be found
by taking the sum of the left- and right-moving photons in a
section of the device. Over one unit cell, one obtains
\be
n_\text{WG} =\left( |{c}_{j}|^2 + |{d}_{j}|^2\right)\frac{d}{c}.
\ee
The relation between this value and the amplitude of the hybrid
Bloch wave may be found by considering the symmetry transformation
used to diagonalize the unit-cell transmission matrix. Defining
${C}_{j}$ to be the amplitude of the Bloch mode of interest, one
finds ${c}_{j}  = s_{11} {C}_{j}$ and ${d}_{j}  = s_{21} {C}_{j}$,
while from the properties of the symmetry matrix $S$, $|{C}_{j}|^2
= |{c}_{j}|^2 + |{d}_{j}|^2$. From here we can deduce the number
of excitations in the waveguide $n_\text{WG}$,  the optical cavity
$n_o$ and the mechanical cavity $n_m$ for a given Bloch wave
amplitude: \bea
n_\text{WG} &=& \frac{d}{c} |{C}_{j}|^2,\\
n_o &=& |{a}|^2\\
&=& \frac{2 |\beta(\delta_k)|^2}{\kappaex} |{c}_{j}  + {d}_{j}|^2\\
&=&   \frac{2 |\beta(\delta_k)|^2}{\kappaex}| s_{11} + s_{21}|^2 |{C}_{j}|^2,\label{eqSI:nofinal}\\
n_m &=& |b|^2 \\ &=& \frac{|\Omega_m|^2}{\delta_k^2 +
\gamma_i^2/4}|{a}|^2. \eea

We then define the fractional occupation in the mechanical mode by
$n_m/(n_\text{WG}+n_0+n_m)$~(with analogous definitions for the
other components). These relations were used to plot the
fractional occupation and colored band diagrams shown in the main
text.

\subsection{Pump Input Power}\label{app:input}

The technicalities associated with pumping the optomechanical
crystal array system are subtly distinct from those in its atomic
system analogue. Due to the periodic nature of the structure and
its strongly coupled property ($\kappaex \gg \kappain$), a bandgap
in the waveguide will arise at the frequency of the ``tuning'' or
pump cavities. This prevents these cavities from being resonantly
pumped via light propagating in the waveguide. This problem may be
circumvented by making the pump frequency off-resonant from
$\omega_2$~(but still at $\omega_1-\omega_m$), and by for example
changing the splitting $\omega_1 - \omega_2$ to be less than the
mechanical frequency. Interestingly, the periodic nature of the
system also allows one to suppress attenuation of the pump beam
through the waveguide~(which would cause inhomogeneous driving of
different elements along the array). Here we calculate the
waveguide input powers required to drive the pump cavities. We
then find the effect of the cavity dissipation rate $\kappain$ on
the beam intensity, to provide estimates for the power drop-off as
a function of distance propagated in the system. Finally we note
that there is a trade-off between required pump intensity and pump
power drop-off. In other words, by tuning the pump beam to a
frequency closer to the pump cavity resonance, the required input
power is reduced, but the attenuation per unit cell
$\exp(-\text{Im}\{K\} d)$ is increased.

To calculate the waveguide input power, we start by considering
the \textit{net} photon flux in the right moving direction, \be
\Phi_R = |{c}_{j}|^2 - |{d}_{j}|^2. \ee Using the properties of
the Bloch transformation matrix $S$ and
equation~(\ref{eqSI:nofinal}), this expression may written in terms
of the number of photons in the optical cavity of interest
($n_{o,j}$), \be \Phi_R = \frac{\kappaex}{2\beta_1^2}
\frac{|s_{11}|^2 - |s_{21}|^2}{|s_{11} + s_{21}|^2}n_{o,j}, \ee
where
\be \beta_1(\delta_k) = \frac{-i\kappaex/2}{-i\kappain/2
-\delta_k}\label{eqSI:beta1}. \ee
The required input power $P_{in}=\hbar \omega_o \Phi_R$ for the
system parameters studied in the main text is shown in
figure~\ref{fig:inputpower}.

To find the attenuation per unit cell $\exp(-\alpha)$, with
$\alpha = \text{Im}\{K(\delta_k)\} d$, of this pump beam, we use a
perturbative approach similar to that used to find the polaritonic
band dispersion. By expanding the Bloch-vector as  $K(\delta_k) =
k^{(0)} + k^{(-1)}/\delta_k + k^{-2}/\delta_k^2 \dots$ and using
equation~(\ref{eqSI:dispersion}), we find $k^{(-1)} = \kappaex/2d$,
and $k^{(-2)} = i \kappaex\kappain/4d$ implying that $\alpha
\approx \kappaex \kappain / 4\delta_k^2$. This approximate
expression is shown along with the exact calculated values for the
attenuation in figure~\ref{fig:inputpower}.

\section{Implementation in an Optomechanical Crystal}

For the theoretical demonstration of our slow-light scheme, we
confine ourselves to a simplified model of an optomechanical
crystal where only the two-dimensional Maxwell equations for TE
waves and the equation of elasticity for in-plane deformations of
a thick slab are taken into account. These equations approximate
fairly accurately the qualitative characteristics of in-plane
optical and mechanical waves in thin slabs, and become exact as
the slab thickness is increased. In this way, many of the
intricacies of the design of high-$Q$ photonic crystal
cavities~(which are treated elsewhere~\cite{safavi-naeini10}) may
be ignored, and the basic design principles can be demonstrated in
a slightly simplified system.

The two-dimensional optomechanical crystal~(2DOMC) system used
here utilizes the ``snowflake'' design~\cite{safavi-naeini10},
which provides large simultaneous photonic and phononic bandgaps
in frequency. Here we choose to use optical wavelengths in the
telecom band, \textit{i.e.}, corresponding to a free-space
wavelength of $\lambda \approx 1.5~\mu\text{m}$. For this
wavelength, we found that the crystal characterized by a lattice
constant $a=400~\text{nm}$, snowflake radius $r = 168~\text{nm}$,
and width $w = 60~\text{nm}$, shown in
figure~\ref{fig:bandstructure}, should work well.

\subsection{Optical and Mechanical Cavities}

\subsubsection{Single Cavity System}

We begin our design by focusing on the creation of a single
optomechanical cavity on the 2DOMC, with one relevant optical and
mechanical mode. This cavity is formed by creating a point defect,
consisting of two adjacent removed holes~(a so-called ``L2''
cavity). We calculate the optical and mechanical spectra of this
cavity using COMSOL, a commercial FEM package, and find a discrete
set of confined modes. Of these, one optical and one mechanical
mode were chosen, exhibiting the most promising value of the
opto-mechanical coupling strength $g$ (see below for calculation).
These modes are shown in figure~\ref{fig:bandstructure}(b) and (c), and
were found to have frequencies $\nu_m = 11.2~\text{GHz}$ and
$\nu_o = 199~\text{THz}$, respectively.

\subsubsection{Double Cavity System}

From here we move to designing the nearly-degenerate double
optical cavity system with large cross-coupling rates. As two
separate L2 cavities are brought close to one another, their
interaction causes the formation of even and odd optical and
mechanical super-modes with splittings in the optical and
mechanical frequencies. This splitting may be tuned by changing
the spacing between the cavities. We take the even and odd optical
modes of this two-cavity system as our optical resonances at
$\omega_2$ and $\omega_1$.

\subsubsection{Optomechanical Coupling Rates}

The optomechanical coupling arises from a shift in the optical
frequency caused by a mechanical deformation. Our Hamiltonian for
the single cavity system can then be written as \be \op{H}{} =
\hbar \omega(\op{x}{}) \opdagger{a}{} \op{a}{} + \hbar \omega_m
\opdagger{b}{}\op{b}{}, \ee where $\op{x}{} = x_{\text{ZPF}}
(\opdagger{b}{} + \op{b}{})$ is the quantized displacement of the
mechanical mode, and $x_{\text{ZPF}}$ is the characteristic
per-phonon displacement amplitude. The deformation-dependent
frequency $\omega(\hat{x})$ may be calculated to first order in
$\hat{x}$ using a variant of the Feynman-Hellman perturbation
theory, the Johnson perturbation theory~\cite{Johnson2002}, which
has been used successfully in the past to model optomechanical
crystal cavities~\cite{eichenfield09,Eichenfield2009a}. The
Hamiltonian is then given to first order by \be \op{H}{} = \hbar
\omega_o \opdagger{a}{}\op{a}{} + \hbar \omega_m
\opdagger{b}{}\op{b}{}+\hbar g (\opdagger{b}{} + \op{b}{})
\opdagger{a}{} \op{a}{} , \ee where $\omega_0$ is the optical mode
frequency in absence of deformation and \be g = \frac{\omega_o}{2}
\sqrt{\frac{\hbar}{2\omega_m d_s}} \frac{ \int dl \left(\m{Q}
\cdot \m{n} \right) \left(\Delta \epsilon |\m{E}^\parallel|^2 -
\Delta(\epsilon^{-1}) |\m{D}^\perp|^2\right)}{\sqrt{\int dA\; \rho
|\m{Q}|^2}\int dA\; \epsilon |\m{E}|^2}. \ee Here $\m{E}$, $\m{D}$
and $\m{Q}$ are the optical mode electric field, optical mode
displacement field and mechanical mode displacement field,
respectively, $d_s$ is the thickness of the slab, and
$\epsilon(r)$ is the dielectric constant.

These concepts can be extended to optically multi-mode systems,
represented by the Hamiltonian \be \op{H}{} =  \hbar \sum_{i}
\omega_{o,i} \opdagger{a}{i}\op{a}{i} + \hbar \omega_m
\opdagger{b}{}\op{b}{}+ \frac{\hbar}{2} \sum_{i,j} g_{i,j}
(\opdagger{b}{} + \op{b}{}) \opdagger{a}{i} \op{a}{j} , \ee where
now the cross-coupling rates can be calculated by the following
expression: \be g_{i,j} = \frac{\omega_{i,j}}{2}
\sqrt{\frac{\hbar}{2\omega_m d_s}} \frac{ \int dl \left(\m{Q}
\cdot \m{n} \right) \left(\Delta \epsilon
\m{E}^{\parallel\ast}_{i}\cdot\m{E}^\parallel_{j} -
\Delta(\epsilon^{-1}) \m{D}^{\perp\ast}_{i}\cdot
\m{D}^\perp_{j}\right)}{\sqrt{\int dA\; \rho |\m{Q}|^2 \int dA\;
\epsilon |\m{E}_i|^2 \int dA \epsilon |\m{E}_j|^2}}. \ee We denote
this expression for convenience as $g_{i,j}\equiv\bra{\m E_i} \m Q
\ket{\m E_j}$.

For the modes of the L2 cavity shown in
figures~\ref{fig:bandstructure}(b) and (c), the optomechanical coupling
was calculated to be $\bra{\m E} \m Q \ket{\m E}/2\pi =
489~\text{kHz}$ for silicon. When two cavities are brought in the
vicinity of each other, super-modes form as shown in
figure~\ref{fig:bandstructure}(a). We denote the symmetric ($+$) and
antisymmetric ($-$) combinations by $\m E_\pm$ and $\m Q_\pm$.
These modes can be written in terms of the modes localized at
cavity 1 and 2, \be \m E_\pm = \frac{\m E_1 \pm \m
E_2}{\sqrt{2}}~\text{and}~\m Q_\pm = \frac{\m Q_1 \pm \m
Q_2}{\sqrt{2}}. \ee

By symmetry, the only non-vanishing coupling term involving the
$\m Q_-$ (antisymmetric mechanical) mode is $\bra{\m E_+} \m Q_-
\ket{\m E_-}$. Assuming that the two cavities are sufficiently
separated, we can approximate $\bra{\m E_+} \m Q_- \ket{\m E_-}
\approx \bra{\m E} \m Q \ket{\m E}/\sqrt{2}$. For the super-modes
of interest, $\bra{\m E_+} \m Q_- \ket{\m E_-} /2\pi = h/2\pi =
346~\text{kHz}$.

\subsection{Properties of Snowflake Crystal Waveguides}
\begin{figure}[htbp]
\begin{center}
\includegraphics[width=17cm]{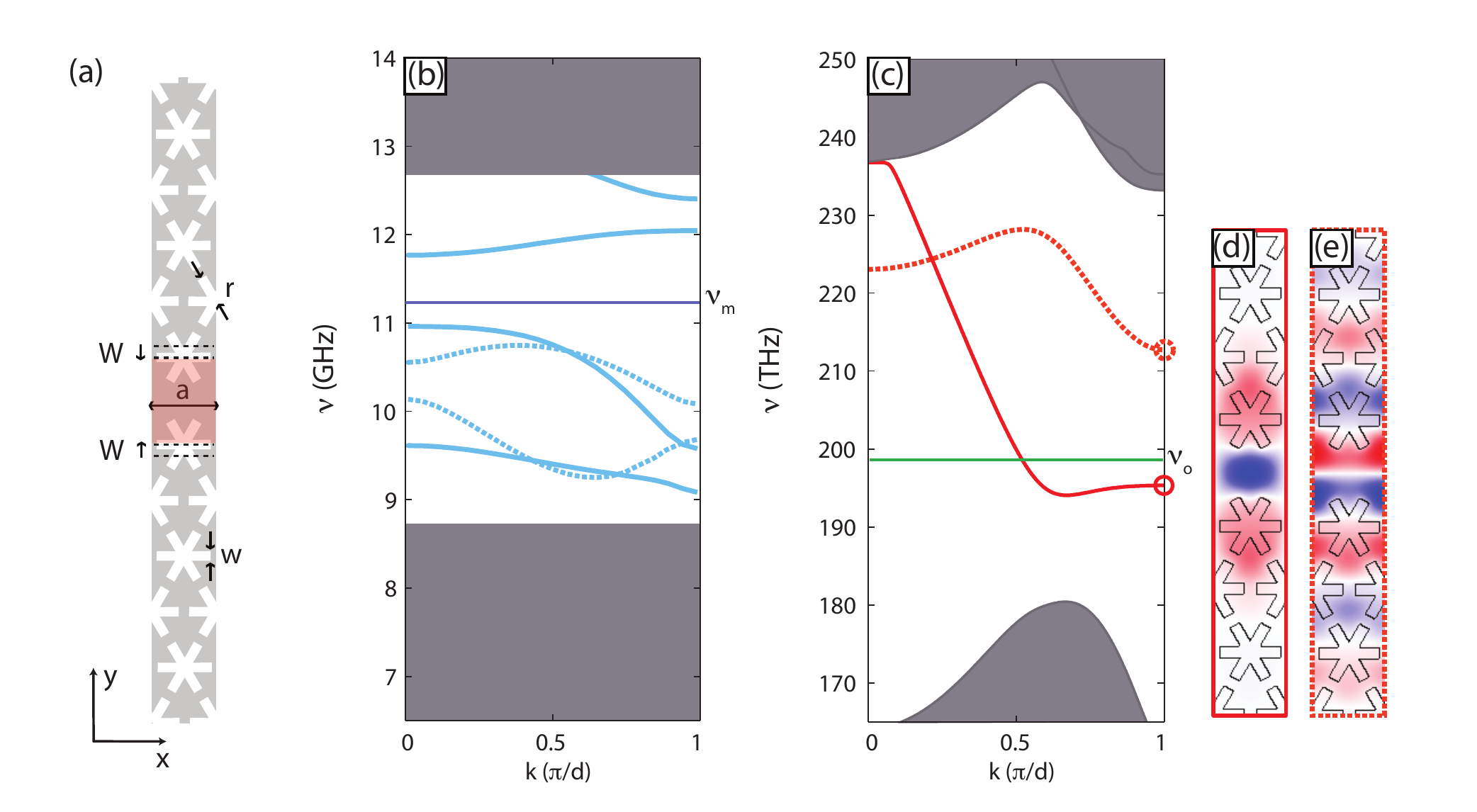}
\end{center}
\caption{(a) The snowflake waveguide consists of the usual
snowflake lattice, with a removed row, and adjusted waveguide
width. The guiding region is shaded red for clarity. (b) The mechanical band structure exhibits a full phononic bandgap at the
resonance frequency of the mechanical mode ($\nu_m$). (c) The optical band structure exhibits a single band that passes through the resonance frequency $\nu_0$ of the optical mode. The waveguide acts as a single-mode optical waveguide with field pattern $H_z$ shown
in (d). The $H_z$ component of the guided optical mode of opposite
symmetry is shown in (e) for completeness.  \label{fig:waveguide}}
\end{figure}

A line defect on an optomechanical crystal acts as a waveguide for
light~\cite{ref:Chutinan2000,ref:Johnson4}. Here, the line defects
used consist of a removed row of holes, with the rows above and
below shifted towards one another by a distance $W$, such that the
distance between the centers of the snowflakes across the line
defect is $\sqrt{3}a - 2W$ (see figure \ref{fig:waveguide}(a)). The
waveguide was designed such that mechanically, it would have no
bands resonant with the cavity frequency (see figure~\ref{fig:waveguide}(b)) and would therefore have no effect on the
mechanical $Q$ factors. Optically, it was designed have a single
band crossing the cavity frequency (see figure~\ref{fig:waveguide}(c))
and would therefore serve as the single-mode optical waveguide
required by the proposal. The band structure of the mechanical
waveguide was calculated using COMSOL~\cite{COMSOL2009}, while for
the optical simulations, MPB~\cite{Johnson2001:mpb} was used.

\subsection{Cavity-Waveguide Coupling}
\begin{figure}[htbp]
\begin{center}
\includegraphics*[width=17cm]{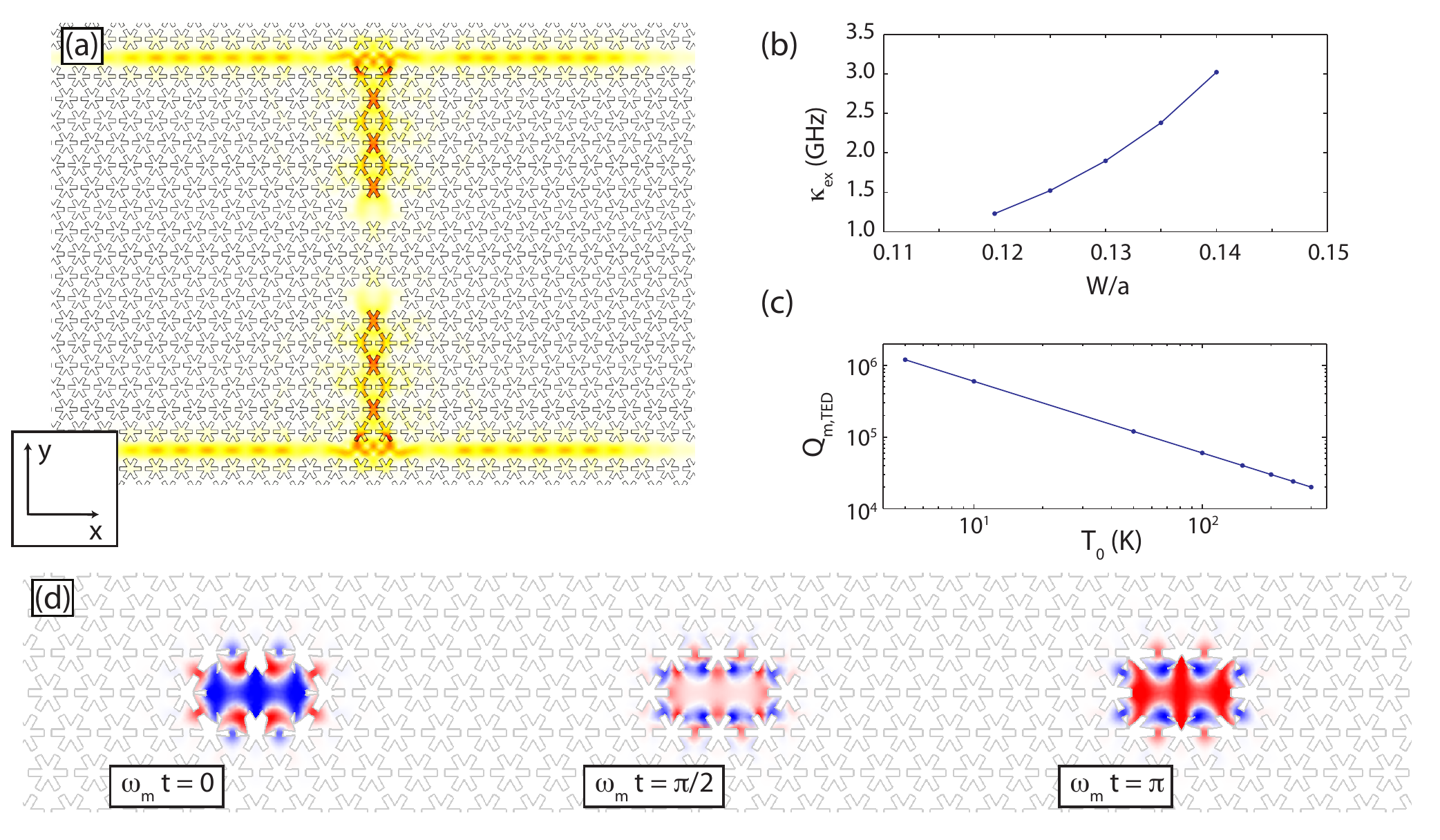}
\end{center}
\caption{(a) A plot of the magnitude of the time-averaged Poynting
vector, $|\avg{\m S}_t|$~(in arbitrary units), shows the leakage
of photons out of the double-cavity system. This induces a loss
rate on the optical modes, which was used to calculate the
extrinsic coupling rate $\kappaex$, plotted in (b) for various values of $W/a$. (c) Plot of the mechanical quality factor $Q_\text{m,TED}$ due to thermoelastic damping, as a function of the ambient temperature $T_0$. (d) The time-harmonic component of the temperature field $\Delta
T(\m r) = T - T_0$ is plotted at various times during a mechanical oscillation period.
\label{fig:coupling_calcs}}
\end{figure}

By bringing the optical waveguide near our cavity, the guided
modes of the line-defect are evanescently coupled to the cavity
mode, and a coupling between the two may be induced, as shown in
figure~\ref{fig:coupling_calcs}(a). Control over this coupling rate is
achieved at a coarse level by changing the distance between the
cavity and waveguide, \textit{i.e.}, the number of unit cells
between them. We found a distance of 6 rows to be sufficient in
placing our coupling rate $\kappaex$ in a desirable range. At this
point, a fine tuning of the coupling rate may be accomplished by
adjustment of the waveguide width parameter $W$, described
previously. The achievable values of $\kappaex$ are plotted
against $W$ in figure~\ref{fig:coupling_calcs}(b). For the final
design, $W = 0.135 a$ is used.

To simulate this coupling rate, we performed finite-element
simulations using COMSOL where we placed the waveguide near our
cavity, and placed absorbing boundaries at the ends of the
waveguide away from the cavity. The resulting time-averaged
Poynting vector $\avg{\m S}_t = |\m E \times \m H^\ast|/2$ is
plotted in figure~\ref{fig:coupling_calcs}(a), showing how the power
flows out of the system.

\subsection{Estimates for Thermoelastic Damping}

The achievable storage times of our system are determined by the lifetimes of the mechanical
resonances. Since we use a phononic crystal, all clamping losses
have been eliminated. However, other fundamental sources of
mechanical dissipation remain, and here we provide estimates for one of these, the
component due to thermoelastic damping
(TED)~\cite{PhysRevB.61.5600,PhysRev.53.90}.

Using the COMSOL finite-element solver~\cite{COMSOL2009}, we
solved the coupled thermal and mechanical equations for this
system~\cite{Duwel2006}. In these simulations the change in the
thermal conductivity and heat capacity of silicon with temperature
were taken into account. The TED-limited quality factors,
$Q_{\text{m,TED}}$ are plotted in figure~\ref{fig:coupling_calcs}(d).
In these simulations, we see that for the mode simulated,
$Q_{\text{m,TED}}$ surpasses $10^6$ at bath temperatures of $T_0 <
5~\text{K}$. To illustrate some representative results of these
simulations, we have plotted the change in
temperature field $\Delta T(\m r)$ from the ambient temperature versus the phase of the mechanical oscillation in figure~\ref{fig:coupling_calcs}(d). At $\omega_m t = \pi/2$,
there are variations in temperature despite the displacement field
$\m Q$ being uniformly $0$ at this time. This shows that at these
frequencies, the temperature does not follow adiabatically the
displacement.

\subsection{Estimate for Optical Pump Heating}

As mentioned in the main text, in a realistic setting the optomechanical driving amplitude $\Omega_m$ itself will be
coupled to the bath temperature through absorption of optical pump photons in the tuning cavities.  This optical pump
heating of the structure is important in estimating the practical limits of the optomechanical system for quantum
applications where thermal noise impacts system performance.  As a realistic model for the bath temperature in our
proposed Silicon optomechanical crystal array, we take $T_b=T_0+\chi\alpha^2$, where $T_0$ is the base temperature and
$\chi$ is a temperature coefficient that describes the temperature rise in each cavity per stored cavity photon due to
optical absorption.  Our estimate of $\chi$ for a thin-film Silicon photonic crystal structure is as follows.  The
absorbed power for $|\alpha|^2$ photons stored in a cavity is given simply by $P_{loss}= \hbar \omega_o \kappa_i
|\alpha|^2$, where $\omega_o$ is the resonance frequency and $\kappa_i$ the optical (intrinsic) linewidth of the cavity.
If we assume \emph{all} of this power is being converted to heat, the change of temperature is $\Delta T = P_{loss}
R_{th}$, where $R_{th}$ is the effective thermal resistance of the Silicon structure.  There are a number of sources in
the literature for $R_{th}$ in relevant photonic crystal geometries~\cite{ref:Barclay7,ref:Notomi09}.  We choose here to
use the value for a two-dimensional crystal system in Silicon, $R_{th} \approx 2.7\times 10^4$~K/W, which yields a per photon
temperature rise of $\chi{\sim}2\;\mu$K assuming an intrinsic loss rate of $\kappa_i\approx 4\times 10^9$~rad/s ($Q_i\approx 3\times 10^6$).


\end{document}